\documentclass[aps,prb,final,twocolumn,showpacs,superscriptaddress]{revtex4}

\usepackage{graphicx}
\usepackage{dcolumn}

\begin{document} 
 
\title{ The role of structural evolution on the quantum conductance behavior of gold nanowires during stretching} 
\author{L.G.C. Rego}
\email[Author to whom correspondence should be addressed: ]{luis.rego@yale.edu}
\altaffiliation[Present address: ]{Chemistry Department, Yale University, 06526, New Haven, CT, USA}
\affiliation{Laborat\'{o}rio Nacional de Luz S\'{\i}ncrotron, Caixa Postal 6192, 13084-971 Campinas SP, Brazil}
\author{A.R. Rocha}
\affiliation{Laborat\'{o}rio Nacional de Luz S\'{\i}ncrotron, Caixa Postal 6192, 13084-971 Campinas SP, Brazil}
\affiliation{Instituto de F\'{\i}sica 'Gleb Wataghin', Caixa Postal 6165, 13083-970, Campinas, SP, Brazil}
\author{V. Rodrigues}
\affiliation{Laborat\'{o}rio Nacional de Luz S\'{\i}ncrotron, Caixa Postal 6192, 13084-971 Campinas SP, Brazil}
\affiliation{Instituto de F\'{\i}sica 'Gleb Wataghin', Caixa Postal 6165, 13083-970, Campinas, SP, Brazil}
\author{D. Ugarte}
\affiliation{Laborat\'{o}rio Nacional de Luz S\'{\i}ncrotron, Caixa Postal 6192, 13084-971 Campinas SP, Brazil}


\date{\today} 
 
\begin{abstract}
Gold nanowires generated by mechanical stretching have been shown to adopt only three kinds of configurations where 
their atomic arrangements adjust such that either the [100], [111] or [110] zone axes lie parallel to the elongation direction.  
We have analyzed the relationship between structural rearrangements and electronic transport behavior during the
 elongation of Au nanowires for each of the three possibilities. We have used two independent experiments to tackle this 
problem, high resolution transmission high resolution electron microscopy to observe the atomic structure and a 
mechanically controlled break junction to measure the transport properties. We have estimated the conductance of nanowires 
using a theoretical method based on the extended H\"uckel theory that takes into account the atom species and their 
positions. Aided by these calculations, we have consistently connected both sets of experimental results and modeled the 
evolution process of gold nanowires whose conductance lies within the first and third conductance quanta. We have also 
presented evidence that carbon acts as a contaminant, lowering the conductance of one-atom-thick wires.
\end{abstract} 
 
\maketitle 
 
\section{Introduction}\label{intro} 

The electron transport through nanometric conductors attract a huge interest due to constant shrinkage of microelectronic devices.\cite{QPC} In 
particular, metal nanowires (NW) display interesting quantum conductance behavior even at room temperature.\cite{MCB} From a practical point of view,
NW's can be easily generated by putting in contact two metal surfaces, which are subsequently pulled apart. During the NW elongation and 
just before rupture, the conductance displays flat plateaus and abrupt jumps, which for metals such as Au, take a value of approximately
one conductance quantum $G_0$ = 2$e^2/h$ (where $e$ is the electron charge and $h$ is Planck's constant).

In spite of the simplicity of the experimental procedure, a new structure with a different evolution is observed for each NW generation and 
all conductance curves have plateaus and jumps, but they display disparate profiles.\cite{MCB} In order to overcome this difficulty, a simple 
statistical method has been usually applied to analyze the average behavior. Instead of considering the conductance as a function of the 
elongation, the transport properties can be represented as a histogram of conductance occurrence, in such a way that a flat plateau 
generates a histogram peak.\cite{MCB} By linearly adding the histograms associated to each conductance curve, a global histogram is generated, which 
describes the general tendencies of an ensemble of NW's. The global histogram displays well defined peaks close to 
the integer multiples of the conductance quantum; this fact has been adopted as a proof of the tendency to conductance quantization in 
metal NW's.\cite{MCB} 

The statistical method, discussed above, provides information on the average behavior but it hinders the study of 
NW transport properties. For example, it is not possible to get the detailed information on how structural 
factors influence the conductance evolution. For several years, the structure evolution was derived from molecular dynamics 
simulations, where the high computer cost imposes the use of simplified potentials based on effective medium theory; 
\cite{Landman1,RefOlesen,Pascual,jellium1,Todorov,Bratkovsky, Landman2,Bahn} subsequently free electron methods were applied to estimate the conduction 
of the metallic neck (or confining potential). More precise methods, considering the electronic structure, have also been applied to 
calculate the NW conductance, but for static atomic configurations.\cite{Landman} 

Recently, {\it in situ} high resolution transmission electron microscopy (HRTEM) experiments have provided a new insight 
in the field. For example, Rodrigues {\it et al.} \cite{Varlei1} have showed that just 
before rupture, gold NW's are crystalline and free of defects and they assume only three kinds of atomic arrangements: two of them form 
bipyramidal constrictions which evolve to one-atom-thick contacts, while the other one generates rod-like NW's that break when they are 
rather thick (three-to-four atoms). By considering that Onishi {\it et al.}\cite{Ohnishi} have already shown that atom size contacts display a 
conductance of 
1 $G_0$, it is possible to discriminate between the rod-like and pyramidal NW morphologies. Further, the relationship between each NW type 
and electrical transport measurements was obtained by simple crystallographic arguments. Validation
of the initial assumption was obtained by statistically comparing the occurrence of observed curve profiles. 

Although these important developments mostly focused on the last conductance plateau, a quantitative understanding of the correlation 
between atomic structure and conductance during the nanowire stretching is still lacking. In this work, we have addressed the 
connection between gold NW structure and the quantized conductance behavior during the NW elongation. 
We have used HRTEM to obtain the detailed information of the atomic structure evolution 
of gold NW's during stretching, and using crystallographic arguments, we proposed the three dimensional structure of these 
nanostructures. 
The NW conductance was measured using an independent, dedicated experimental set-up: a mechanically controlled
break junction operated in ultra-high-vacuum (UHV-MCBJ). 
In order to correlate both sets of data, we have used a  semi-empirical atomistic theoretical technique based on the extended
H\"uckel theory\cite{Glynn} (EHT) that allows for the estimation of transport properties of systems with a large number of atoms.\cite{Kirczenow}
The results displayed an excellent agreement between observed structural and conductance experiments and theoretical calculations, 
enabling a complete modeling of the NW elongation process both from the structural and transport properties points of view.

The remaining of this work is organized as follows: section \ref{experiment} describes the experimental techniques used to obtain the 
structural and transport data, respectively; section \ref{theory} describes the theoretical model employed to calculate the electronic 
quantum transport through the metallic NW's; the experimental results are presented in section \ref{expresul} and the analysis of the data 
via theoretical analysis is presented in sections \ref{discuss1} 
for NW's oriented along the [100], [111] and [110] directions as well as \ref{discuss2},
where the presence of impurities is investigated; finally, in section \ref{conclu} we draw our conclusions. 

\section{Experimental Apparatus}\label{experiment}

We have generated NW's \textit{in situ} in a HRTEM (JEM 3010 UHR, operated at 300 kV, point resolution 1.7 {\AA}) using the method developed 
by Takayanagi's group.\cite{Ohnishi,RefKondo} The procedure consists in focusing the microscope electron beam 
(current density 120 A/cm$^{2}$) on the sample to perforate and grow neighboring holes until a nanometric bridge 
is formed between two of them. We have used a self-supported polycrystalline gold thin film (5 nm thick, 
deposited on a holey carbon grid) in order to generate NW's between apexes of different orientations and elongate 
them in different directions.\cite{Varlei1} Atomic resolution image acquisition has been performed after reducing 
the electron beam intensity to its usual value (30 A/cm$^{2}$). A high sensitivity TV camera (Gatan 622SC, 30 frames/s) associated with a 
conventional video recorder was used to register NW real-time evolution. The procedure described above, allows us to generate NW's with a 
remarkable stability, because the NW, its apexes and the surrounding thin film regions, form a monolithic block. In consequence, the NW's 
formed by a few atomic layers usually show a long life time (1-10 minutes). Despite this stability, the generated NW's elongate 
spontaneously, get thinner and then break due to the relative slow movement of the NW apexes. These apexes 
displacement are probably due to a film deformation induced by thermal gradients between parts of the sample, 
as usually observed in TEM thin film work. A critical aspect when studying such tiny nanostructures is the 
presence of contaminants, the most critical one in TEM is amorphous carbon.\cite{reimer} However, we must keep in mind, 
that the intense electron irradiation transforms carbon into bucky-onions and finally clean the gold 
surface.\cite{RefUgarte} 

The electric transport properties of gold NW's were studied using an independent instrument specially designed 
for this purpose, namely an UHV-MCBJ.\cite{RefMCBJ} In this approach, a macroscopic gold wire 
(99.99 $\%$ pure, $\phi$  = 75 $\mu$m) is glued in a flexible substrate in two points; then it is weakened
in a point between the two fixing parts by an incomplete cut. By bending the substrate \textit{in situ} in 
the UHV, we break the wire and produce two clean gold surfaces; using the same bending movement, the fresh 
tips are put together and separated repeatedly in order to generated and deform NW's. It's important to remark 
that in this configuration the NW's are generated from surfaces obtained in UHV ($<$ 10$^{-8}$ Pa), so it is 
expected to have a clean sample for few hours.\cite{RefMCBJ} This care with vacuum conditions is extremely 
important to generate reliable experimental measurements on NW properties.\cite{Varlei1,RefHansen}

Electrical measurements were done using a home-made voltage source, and current-voltage converter powered with 
isolated batteries to reduce electrical noise. Data acquisition was based on an eight-bit digital oscilloscope 
(Tektronic TDS540C, 2 GSample/s, bandwidth 500 MHz); this electronic system was developed with the aim of 
improving the conductance measurement precision, yielding values with a relative error of 
($\Delta G/G$) $\sim$ 10$^{-4}$. The applied voltage was 100 mV, and the conductance was measured in the [0, 2.8] $G_{0}$ range to improve 
the detection of the last two quantum conductance plateaus.  
 
\section{Theory}\label{theory} 
 
We follow the approach introduced by Emberly and Kirczenow \cite{Kirczenow} that is based on the extended H\"uckel theory, which is a
method traditionally used in quantum chemistry to calculate the electronic properties of clusters and molecules.\cite{Glynn,YAeHMOP} One 
advantage of this approach lies in the possibility of treating the whole nanocontact -- both tips and the nanowire between them -- as a 
unique entity, described by molecular orbitals. 
 
The tips and the nanowire, upon which we base our transport calculations, are built on the basis of the crystallographic properties of the
material as well as the obtained HRTEM images; no attempt was made to simulate the dynamics of the NW formation. In our calculations the 
nanocontacts are made up of as many as 150 gold atoms for some structures.  In order to calculate the quantum electronic states of Au 
clusters as big as these, a semi-empirical method is necessary to produce results within reasonable time. For this reason the EHT 
is employed; the method includes overlap terms and matrix elements that are not restricted to first neighbors; 
moreover, all the valence states -- 5d, 6s and 6p, in the case of Au -- are considered. One drawback is the fact that, in its standard form,
the EHT is a one-electron formalism that does not take into account explicitly electron-electron interaction effects. However, we believe
that this problem can be partially circumvented, because of the parameterization of the orbital wavefunctions by self-consistent 
Hartree-Fock states and  the use of semi-empirical energy parameters. Such effects are more relevant for the analysis of the structural 
stability and relaxation of nanowires, where the calculation of energy manifolds is necessary. Nonetheless, tight-binding together with 
molecular dynamics methods, similar to the present one,  have been successfully used to simulate the dynamics of the nanowire 
formation.\cite{Zacarias}

\begin{figure}[h]
\includegraphics[width=6.5cm]{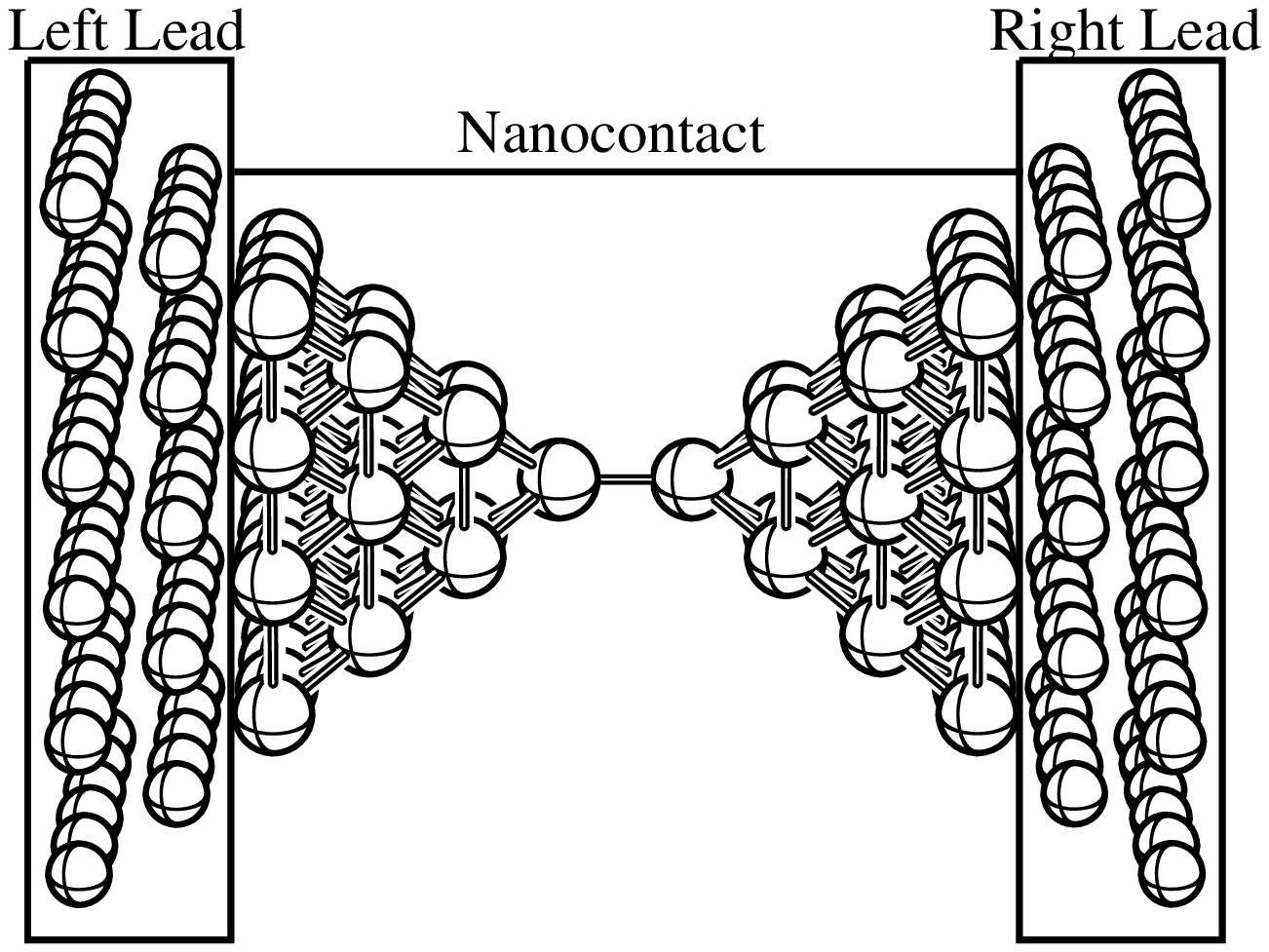}
\caption{Sketch of model structure used in the calculations evidencing one unit-cell of the incoming and outgoing leads and the actual
NW.}
\label{drawing}
\end{figure}

Let us assume that a nanocontact is formed by two apexes (or tips) and the nanowire itself that connects them. We start the 
description by defining the molecular Hamiltonian of the nanocontact, $H_{NC}$ and its molecular orbitals (MO's) $|\phi\rangle$: 
\begin{eqnarray} 
|\phi\rangle &=& \sum_{\nu} \sum_j^{NC} A_{\phi,\nu j} |\nu j\rangle \ ,  
\label{MO} 
\\ 
\epsilon_{\phi} &=& \langle \phi|H_{NC}|\phi\rangle.  
\label{E_NC} 
\end{eqnarray} 
In (\ref{MO}) the sum over $\nu$ extends over the $5d$, $6s$ and $6p$ orbitals and the index $j$ covers all the Au atoms in the nanocontact. The 
molecular orbitals $|\phi\rangle$, written as a linear combination of atomic orbitals in (\ref{MO}), obey the orthonormality relation  
$\langle \phi'|\phi \rangle = \delta_{\phi'\phi}$. Equation (\ref{E_NC}) defines the MO energies. 
So that these MO's bear quantum phase coherence along the entire nanocontact.  
 
To account for electron transport the nanocontact is coupled to infinite Au leads on both sides (left (${\cal L}$) and 
right (${\cal R}$) of the nanocontact), which behave as charge reservoirs (see Fig. 
\ref{drawing}). The leads are divided into unit cells whose MO's are also obtained by the EHT method: 
\begin{eqnarray} 
|\alpha\rangle &=& \sum_j^{cell} B_{\alpha,j} |s,j\rangle \ , 
\label{Bloch} 
\\ 
\epsilon_{\alpha}^{\cal{L}(\cal{R})} &=& \langle \alpha | H_{\cal{L}(\cal{R})} | \alpha \rangle \ . 
\label{E_LR} 
\end{eqnarray} 
For the sake of simplicity only $6s$ Au orbitals are used in (\ref{Bloch}). The Hamiltonian 
$H_{\cal{L}(\cal{R})}$ includes overlap terms and is not restricted to first neighbor matrix elements within the unit cell. We consider
this approximation because the contacts are expected to behave as bulk gold, hence s orbitals are adequate for describing its electronic 
properties. In the nanoconstriction, where the number of atoms is smaller, we expect a different behavior. Although 
$\langle \alpha' | \alpha \rangle = \delta_{\alpha'\alpha}$, it is noteworthy that the scalar product between both types of MO's define the
overlap term: 
\begin{eqnarray} 
\langle \alpha | \phi \rangle = S_{\alpha\phi} \ . 
\label{S} 
\end{eqnarray}  
 
Having defined the basic physical quantities, let us consider the total Hamiltonian  
\begin{eqnarray} 
H = H_{\cal{L}} + H_{\cal{NC}} + H_{\cal{R}} + W_{\cal{L}} + W_{\cal{R}} \ . 
\label{H} 
\end{eqnarray} 
The term $W_{\cal{L}}$ in (\ref{H}) is responsible for coupling  the nanocontact with the first unit cell of the lead on its left and 
$W_{\cal{R}}$, similarly, with respect to the first unit cell on the right. In principle $H_{\cal{L}}$ and $H_{\cal{R}}$ can be 
different from each  other, as in the case of tips with distinct crystallographic orientations or defects. 

To account for the transport calculations consider Bloch electrons that propagate through the leads:
\begin{eqnarray}
|\Psi \rangle = \sum_n e^{i n \theta_{\beta}} |\beta\ n\rangle \ ,
\label{Bloch1}
\end{eqnarray}
with $|\beta\ n \rangle$ as one of the MO's of the unit cell at position $n$ and $|\beta \rangle \in \{|\alpha \rangle \}$. At the site of 
the nanocontact the incoming Bloch wave is scattered giving rise to reflected and transmitted waves, according to the framework of the multi-channel
Landauer formalism.\cite{Ferry} Therefore, for a rightward incoming  electron we write the following wavefunction 
$|\Psi\rangle = |\Psi_{\cal{L}}\rangle + |\Psi_{\cal{NC}}\rangle + |\Psi_{\cal{R}}\rangle$, where 
\begin{eqnarray} 
|\Psi_{\cal{L}}\rangle \ \ &=& \sum_{n<0} \left\{ e^{in\theta_{\beta}} |\beta\ n\rangle +  
\sum_{\alpha \in \cal{L}} r_{\alpha \beta} e^{-in\theta_{\alpha}} |\alpha\ n\rangle \right\},  
\label{Psi_L} 
\\  
|\Psi_{\cal{NC}}\rangle &=& \sum_{\phi} C_{\phi\beta} |\phi\ 0\rangle \ ,
\label{Psi_NC} 
 \\ 
|\Psi_{\cal{R}} \rangle \ \ &=& \sum_{n>0} \sum_{\alpha \in \cal{R}} t_{\alpha\beta} e^{in\theta_{\alpha}} |\alpha\ n\rangle \ . 
\label{Psi_R} 
\label{wf} 
\end{eqnarray}   
The incoming electron has energy $E(\theta_{\beta}) \ge \epsilon^{\cal{L}}_{\beta}$ in order to be a propagating state. In expression 
(\ref{Psi_NC}), $|\phi\ 0\rangle$ represents the $\phi$-MO's of the nanocontact, which occupies the site $n=0$, and $C_{\phi\beta}$ are 
unknown coefficients. Similarly, in (\ref{Psi_L}) and (\ref{Psi_R}), $|\alpha\ n\rangle$ is the $\alpha$-MO of the unit cell at position $n$;  
$r_{\alpha\beta}$ and $t_{\alpha\beta}$ are the channel reflection and transmission coefficients, to be determined in the sequence, which
will be used to calculate the total transmission $T(E)$ and reflection $R(E)$ of electrons propagating through the nanocontact. The parameter
$\theta$ is a dimensionless wave-vector for the Bloch waves in the leads. 
 
In order to determine the coefficients $\{r_{\alpha \beta},t_{\alpha \beta}, C_{\phi \alpha}\}$ a complete set of MO's  
$\{\xi\} \equiv \{\phi\} \cup \{\alpha\}$ is put together, where we project the equation $H |\Psi \rangle = E |\Psi \rangle$. As a result
we obtain a non-homogeneous system of coupled equations that can be solved numerically to yield these coefficients. A detailed description 
is provided in the appendix.

Concluding the procedure, we define the total transmission and reflection coefficients, which are written in terms of the incoming and  
outgoing coefficients and the electron velocity $v_{\alpha}$ of the respective Bloch wave mode:  
\begin{eqnarray} 
T_{\beta}(E) &=& {\sum_{\alpha \in \cal{R}}}' \ \ \frac{v_{\alpha}}{v_{\beta}} |t_{\alpha \beta}|^2 \ , 
\label{T} 
\\ 
R_{\beta}(E) &=& {\sum_{\alpha \in \cal{L}}}' \ \ \frac{v_{\alpha}}{v_{\beta}} |r_{\alpha \beta}|^2 \ , 
\label{R} 
\end{eqnarray} 
where the prime means that the sum is made exclusively over the propagating modes, that is, $E(\theta_{\alpha}) \ge \epsilon_{\alpha}$. 

Finally, the Landauer formula \cite{MCB,Landauer} is used to determine the charge transport through the nanocontact. For a degenerate metallic system in the 
linear response regime, the electrochemical potential difference between source and drain takes the limit $\Delta \mu \rightarrow 0$ and the 
conductance $G= I / V$ becomes 
\begin{eqnarray} 
G = \frac{2 e^2}{h} {\sum_{\beta}}' \int dE \ T_{\beta}(E) \ \frac{\partial \eta}{\partial \mu} =  
\frac{2 e^2}{h} \ {\sum_{\beta}}' T_{\beta}(E_F)\ , \  
\label{G} 
\end{eqnarray} 
where $\eta$ is the Fermi-Dirac distribution and $E_F$ is the Fermi energy. Since the voltage applied across the leads is small compared to 
$E_F$ the conductance is to be considered when the electron energy is close to this value.
 
\section{Experimental Results}\label{expresul}

\begin{figure}[h]
\includegraphics[width=8.5cm]{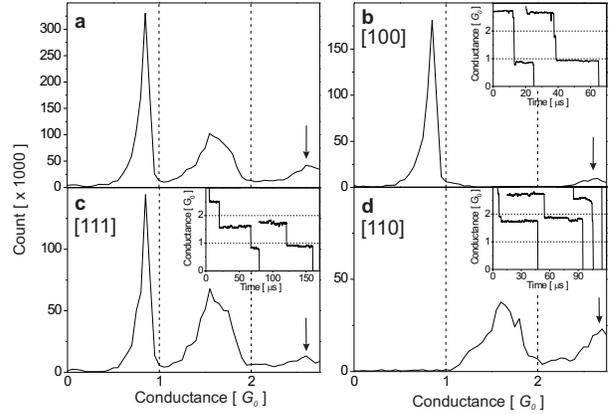}
\caption{(a) Au NW global histogram and (b-d) histograms associated with each of the three types of curve profile (see text for explanation).
The greatest recorded value shown in the histogram's abscissa is 2.8 $G_0$. The arrows show the existence of peaks below 3 $G_0$. The insets
of every figure show typical conductance curves for each direction.}\label{hist}
\end{figure}

Figure \ref{hist}(a) shows the global histogram generated from a series of 500 collected conductance curves of gold NW's measured using the 
UHV-MCBJ. Three peaks can be identified in the histogram; as expected the major one is close to 1 $G_0$ and, the two minor ones are at 
1.8 and 2.8 $G_0$. 

Rodrigues {\it et al.}\cite{Varlei1} have demonstrated that analyzing the occurrence of the last two conductance plateaus (around 1 and 2 
$G_0$), the conductance curves can be grouped in three different classes: a) only a plateau at 1 $G_0$ b) plateaus at 1 and 2 $G_0$; and, finally, 
c) no plateau at 1 $G_0$. These characteristics are associated with the crystallographic direction along which the NW is being 
elongated, corresponding to the [100], [111] and [110], respectively. As the global histogram superposes and hides this structural 
information, we have built three histograms formed by each group of conductance curves (Fig. \ref{hist}(b-d)), examples of the conductance curves 
are presented at the insets). This approach allows us to obtain a statistical validation for the sequence of conductance plateaus 
appearing during the NW elongation along each axis, extending the range from 2 to 3 $G_0$. In these terms, it can be deduced that the 
thinning of a NW along [100] axis (hereafter denoted [100] NW) generates a conductance curve with plateaus close to 3 and 1 $G_0$
(see curves inset Fig. 2(b)); a [111] NW, plateaus around 3, 2 and 1 $G_0$ and; finally, for [110] NW's, plateaus around 3 and 2 $G_0$. 

\begin{figure}[h] 
\includegraphics[width=8.5cm]{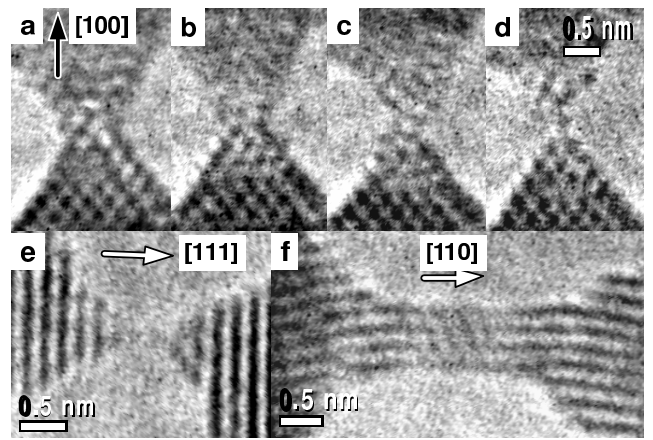} 
\caption{Real time HRETM images showing typical nanowires formed in the direction of the three zone axes. (a-d) Elongation process of a 
[100]-oriented NW; note the existence of two misaligned planes just before the formation of the ATC. (e) ATC oriented along the [111] 
direction and (f) rod-like structure formed along the [110] direction.}\label{HRTEM} 
\end{figure} 

Figure \ref{HRTEM} shows typical HRTEM micrographs of the three possible Au NW structures; a common feature of all the observed NW's is the fact 
that they are defect free crystals. Figures \ref{HRTEM}(a-d) show snapshots of the elongation process of a NW along the [100] direction, 
where the progressive thinning of the constriction can be followed with atomic resolution. We can see from these images that the NW 
stretches from a symmetrical bipyramidal NW with a two-atom wide minimal section (Fig. \ref{HRTEM}(b)) to a constriction formed by two 
misaligned truncated pyramids with a two atom wide top (Fig. \ref{HRTEM}(c)). Finally, a one-atom-thick contact is formed 
(Fig. \ref{HRTEM}(d)); this contact subsequently evolves to form a suspended chain of atoms (ATC) before rupture (not shown). Figure 
\ref{HRTEM}(e) allows the clear visualization of the bipyramidal morphology of a one-atom-thick contact formed for [111] NW's. In contrast, 
[110] NW's generate rod-like structures before rupture (Fig. \ref{HRTEM}(f)). 

At this point, it is useful to remind some basic aspects of 
the reported {\it in situ} HRTEM experiments. Firstly, HRTEM images of NW's can be described as a two-dimensional projection of the atomic 
potential, 
then good quality atomic resolution information is only obtained from crystals very well oriented in relation to the microscope electron 
beam. In our experiments, we can not easily control the orientation of the apexes forming the nanometric constrictions, so the possibility 
of getting sequences as shown in Fig. \ref{HRTEM}(a-d) ([100] NW) is statistical and unpredictable. 

Unfortunately, this kind of detailed 
information on the atomic arrangement evolution could not be observed for the other two NW types of nanowires ([111] and [110]). Another 
limiting 
factor of our dynamical HRTEM observations is the temporal sampling of our image acquisition system allowing the capture of 30 frames per 
second; very rapid events ($<$ 1/30 s) may go by undetected. 

Despite these difficulties, HRTEM images provide a wealth of remarkable information to derive the atomic structure of gold NW's from. By 
combining the atomic resolution micrographs (two dimensional projections of the structure) and the known surface energy properties of the 
analyzed systems it is possible to deduce the three dimensional atomic arrangement of NW's. This procedure is based on the so-called Wulff 
construction that allows a quick geometrical determination of the faceting pattern of nanostructures;\cite{Marks} such approach has 
already been successfully applied to study Au,\cite{Varlei1} Ag\cite{LNLS} and Pt\cite{RefPRLPt} NW's. 

In the following sections, we present a careful and thorough analysis of the 
derived atomic arrangements and their evolution for the three kinds of Au NW's; the theoretically obtained electronic transport properties 
of such structures are compared with the conductance behavior observed with the UHV-MCBJ experiment.

\section{Discussion: Correlation between atomic structure and electronic effects}\label{discuss1}

\subsection{[100] Gold NW's}\label{[100]}

\begin{figure}[ht]
\includegraphics[width=7.5cm]{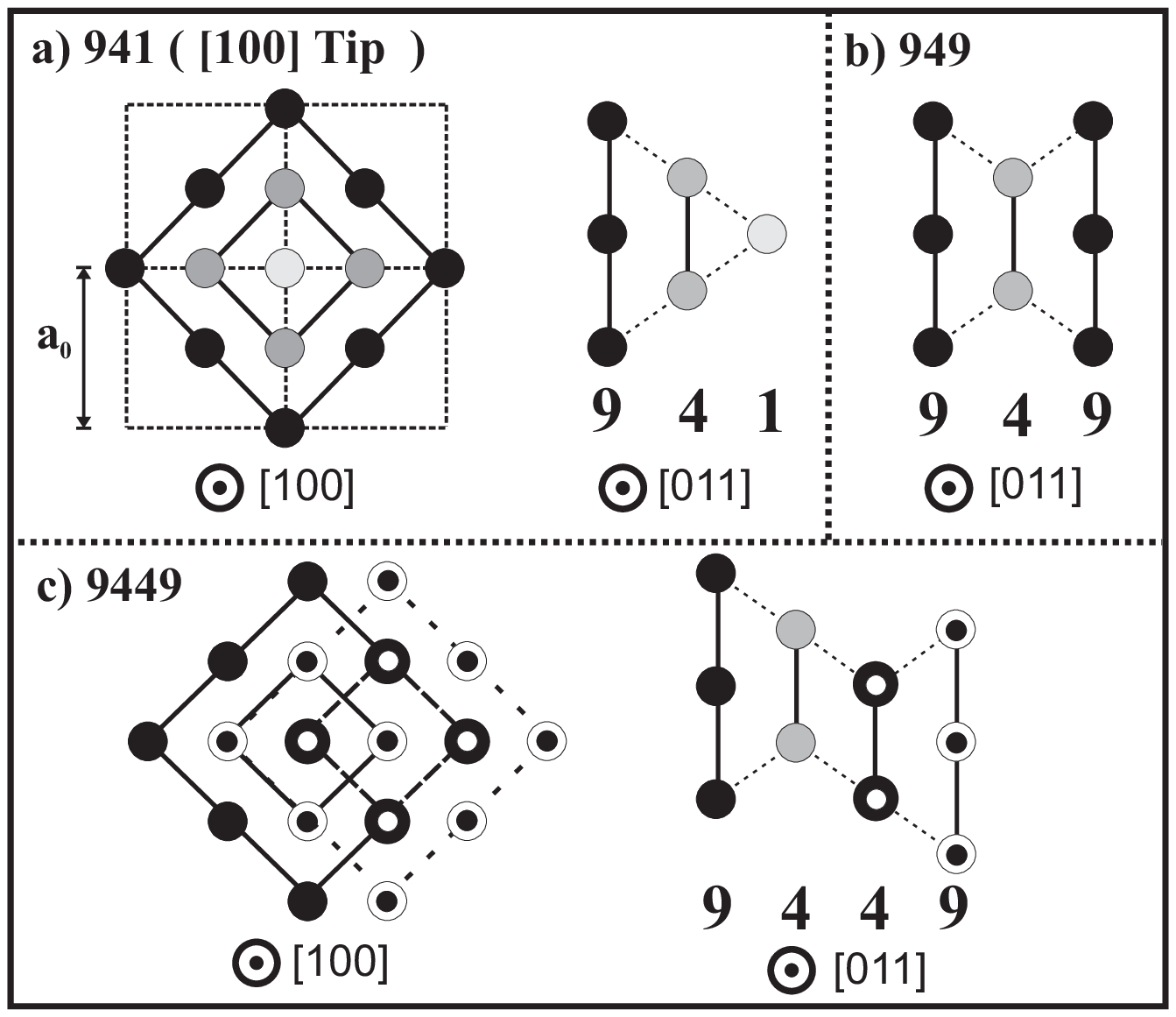}
\caption{Atomic geometry of different Au nanocontacts oriented along the [100] direction. 
The notation indicates the number of atoms in the planes perpendicular to the [100] axis in the constriction region:
(a) 941, (b) 949 and (c) 9449 tips. The dashed lines in (a) indicate the edge of the fcc unit-cell and a$_0$ is the lattice parameter.}
\label{100}
\end{figure}
 
The [100] nanocontacts are generated by two pyramidal tips, each of them faceted by four low energy (111) planes; they can be elongated 
until atomic-chains are formed, having generally two-to-four atoms in length.\cite{Varlei1} Figure \ref{100} illustrates three nanocontact 
geometries based on the crystallographic structure of face-centered-cubic (fcc) bulk gold and the Wulff method. These structures simulate the steps of the NW 
evolution in the experiments. Figure \ref{100}(a) shows the cross-section ([100] direction) and side view ([011] direction) of an atomically
sharp [100] NW tip; the side view presented represents the structural arrangement as seen from HRTEM images in Fig. \ref{HRTEM}(a-d). 
This particular tip is formed by the stacking of a plane of 9 atoms (black circles) followed by a plane of 4 atoms 
(dark gray circles) and, finally, one atom at the end of the tip (light gray circle). The notation used hereafter to designate the 
structures is given by the number of atoms that make up the alternate planes stacked along the NW orientation axis. The one-atom-thick 
contact (941149) which is the structure observed just before rupture (example is presented in Fig. \ref{HRTEM}(e)) 
is then formed by mirror imaging the tip arrangement described above. We can use this pyramidal tip arrangement to model the thicker NW's
observed prior to the formation of the 941149 structure, {\it i. e.}, we can build thicker structures by removing planes of atoms from
the tip arrangement in Fig. \ref{100}(a). Hence, in simple terms, we would expect that the second thinnest [100] constriction would be 
formed by pyramids where the one atom tip is removed leading to a minimum cross-section of four atoms. By joining two pyramids sharing the 
four-atom plane we generate the 949 neck shown in Fig. \ref{100}(b) and corresponding to the experimental results in Fig. \ref{HRTEM}(b).

However, HRTEM data clearly show the existence of an intermediate structure (Fig. \ref{HRTEM}(c)) where the two truncated pyramids forming 
the apexes are not aligned with respect to each other. Assuming that all the neck region is defect free ({\it i. e.} a perfect arrangement
of atoms in a fcc lattice) and following the Wullf morphology we must conclude that the NW is formed by two four-atom planes which are not 
not aligned with each other, rather the corner of one of these planes is located at the central axis of the other and vice-versa. This 
structure is presented in Fig. \ref{100}(c) and labeled 9449.

\begin{figure}[h]
\includegraphics[width=6.5cm,clip=true]{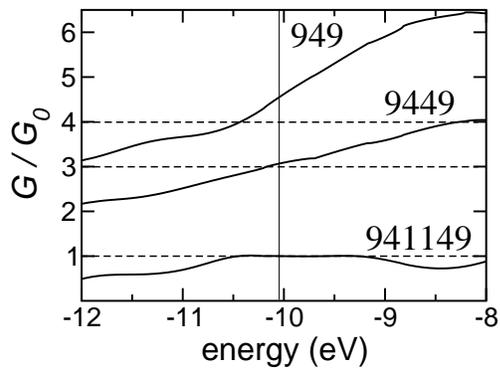}
\caption{Calculations of the quantum conductance ($G$) as a function of the electron energy for various 
NW structures oriented along the [100] direction: 941149 (the one-atom-thick contact), the 9449 and 949 nanocontacts. 
The vertical line indicates the Fermi energy; the conductance is in units of the quantum $G_0$.}  
\label{theo-100}
\end{figure} 

Using the formalism of section \ref{theory}, we have calculated the quantum conductance for the NW morphologies shown in Fig. 
\ref{100} as a function of the electron energy. The results are shown in Fig. \ref{theo-100}, where the vertical line indicates the value 
of the Fermi energy in the leads. At first, we notice that the one-atom-thick contact (941149, distance between the atoms is 2.88 {\AA} 
even at the contact) shows a conductance curve that yields a very stable plateau at $G(E_F)$ = 1 $G_0$ which corresponds to a peak around 1
$G_0$ in the histograms (Fig. \ref{hist}(b)). On the other hand, the 9449 nanocontact display a conductance of $G(E_F)$ $\sim$ 3 $G_0$, and the 949, 
4.5 $G_0$.
     
In brief, the one-atom-thick (941149) and the ATC's conductance (9411149, also calculated but not shown here) is $\sim$ 1 $G_0$ while the 
9449 structure shows a conductance  
of $\sim$ 3 $G_0$. Each of these conductance values agrees very well with the peaks in the global conductance histogram for the [100] 
zone axis (Fig. \ref{hist}(b)). In other words, he have accounted for the peaks in the histogram corroborating the assumption that the 
sequence of atomic arrangements shown in Fig. \ref{100} can, in fact, represent the dynamics of NW evolution along the [100] direction.
It is also worth noticing that the deduced structural evolution shows no contribution around 2 $G_0$, in agreement with the 
conductance histogram. Finally, the 949 morphology presents $G(E_F)$ $\sim$ 4.5 $G_0$, which would represent a thicker structure however, 
our experiments have not included conductance measurements in that range.    
 
\subsection{[111] Gold NW's}\label{[111]}

\begin{figure}[h]
\includegraphics[width=8.5cm,clip=true]{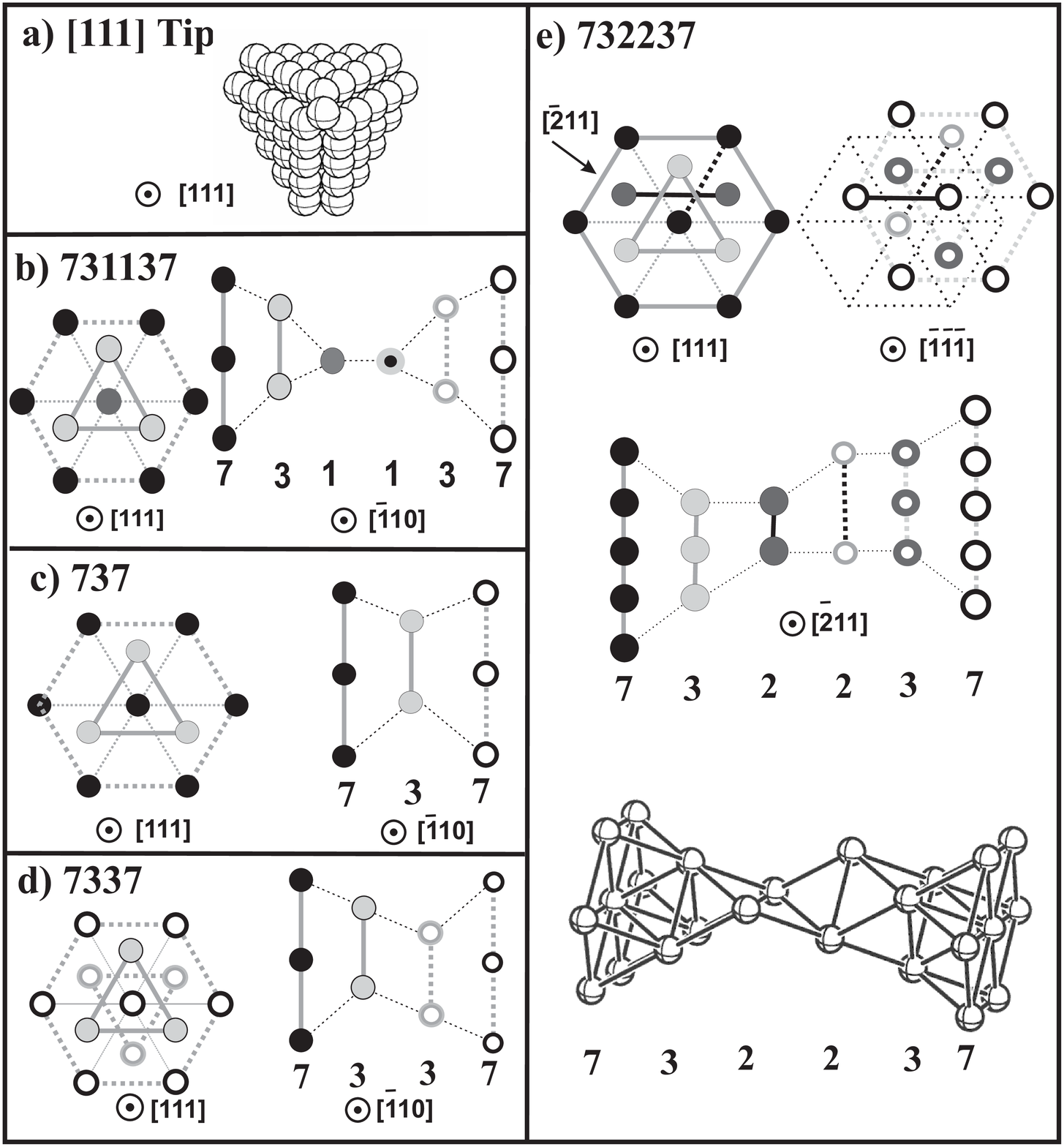}
\caption{Atomic geometry of different Au nanocontacts oriented along the [111] direction. Figure (a) shows the arrangement 
of the tip. The notation indicates the number of atoms in the planes perpendicular to the [111] axis in the constriction region:
(b) 731137, (c) 737, (d) 7337, and (e) 732237}
\label{111}
\end{figure}

For the case of [111] NW's, we have no atomic resolution HRTEM images that show the structural evolution during the thinning process. 
However, we can model the structures using the same procedure of section \ref{[100]},\cite{Varlei1} {\it i. e.}, we can start with the 
one-atom-thick
nanowires and make our way up towards thicker structures. Again, the tip morphology was built from a fcc crystal. It is worth noting that
the stacking of planes in the [111] direction is made by three hexagonal planes placed at alternate intersite spaces (usually labeled 
ABCABC$\ldots$). The derived [111] pyramidal morphology is presented in Fig. \ref{111}(a). This arrangement is obtained from the smallest 
possible fcc gold nanoparticle, {\it i. e.}, the cuboctahedral with regular hexagonal faces based on the Wulff method.
\cite{Varlei1}

As it was done in the previous section, the one-atom-thick NW is formed by two aligned [111] apexes oriented in opposite directions. We 
can label this structure 731137 by following the notation introduced earlier. This structure is depicted in Fig. \ref{111}(b). In analogy
with the procedure adopted for the [100] direction, we would initially expect that the removal of the last atom from the tip and then by 
joining the remaining truncated pyramids we would have the arrangement before the one-atom-thick NW is formed. This particular
arrangement is sketched in Fig. \ref{111}(c) and is named 737. However, this arrangement is defective because it forms a twin at the neck.
This twin should be unfavorable in the sense that the NW's are free of defects in this size range.\cite{Varlei1,Rubio,Durig}. 
In that case, we proposed a structure analogous to
the 9449 [100] NW which was presented in section \ref{[100]}. This new structure is formed by adjoining two truncated [111] pyramids (73) 
rotated 180$^o$ with respect to each other. The resulting arrangement has a perfect fcc stacking as shown in Fig. \ref{111}(d). 
Finally, molecular dynamics simulations using classical potentials\cite{douglas1} indicate the existence of a thinner region with two 
misaligned dimmers. As a matter of fact, this arrangement is formed by simply removing atoms from the apexes of the 7337 NW. This 
structure, labeled 732237, is presented in Fig. \ref{111}(e) for atoms in a perfect fcc arrangement.

\begin{figure}[h]
\center
\includegraphics[width=8.5 cm,clip=true]{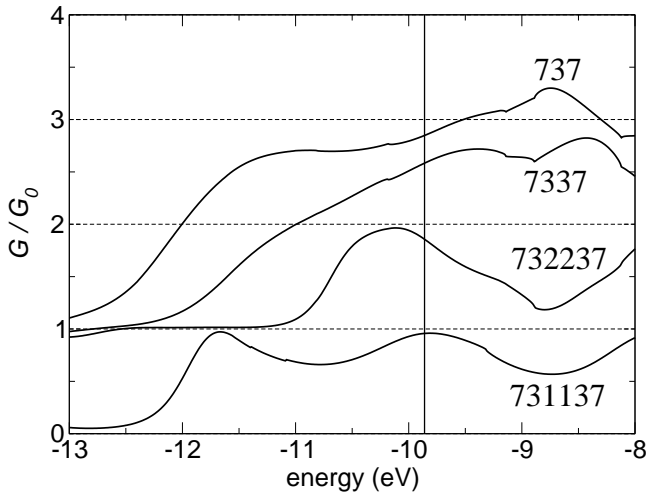}
\caption{ Calculations of the quantum conductance $G$ as a function of the electron energy for various proposed nanocontact structures oriented 
along the [111] direction: the 737, 7337, 732237 and the one-atom-thick (731137) nanocontacts.}
\label{theo-111} 
\end{figure}

In Fig. \ref{theo-111}, we present the theoretical conductance calculations for each of the proposed structures for [111] NW's. As 
previously stated, the experimental results must be compared to the calculated conductance at the Fermi energy level, represented by the
vertical line in the figure. Although the 737 arrangement (Fig. \ref{111}(c)) would be a possible configuration displaying a conductance of
$\sim$ 2.8 $G_0$, we must bear in mind that it presents a twin defect. Hence we have neglected it in our analysis. The calculated 
conductance for the 7337 structure, which presents a perfect fcc stacking, is below 2.6 $G_0$ as shown in Fig. \ref{theo-111}. 
Conductance measurements for this orientation (Fig. \ref{hist}(c)), show the existence of a peak close to 2 $G_0$. In this sense, the 732237 
NW, results in $G(E_F)$ $\sim$ 1.9 $G_0$ as can be seen in Fig. \ref{theo-111}. Finally, structure 731137 yields $G$ = 1 $G_0$ (Fig. 
\ref{theo-111}(d)).

One might be led to believe that the relaxation of the thinnest region (..22..) of the 732237 structure will rotate the two dimmers to form
a perfect tetrahedron in order to maximize the number of neighbors. This deformation generates a NW with two [111] apexes bounding a 
[100]-oriented neck. This deformation decreases the conductance by a significant amount from 1.9 to $\sim$ 1.2 $G_0$; emphasizing that
nanowires are particularly sensitive to breaking symmetries and to defects in the neck region. 

In other words, the 7337 morphology has a conduction close to 2.6 $G_0$, the 732237 NW shows the calculated conductance value $G(E_F)$ 
$\sim$ 1.9 $G_0$ and at last, for the one-atom-thick contact, $G(E_F)$ $\sim$ 1 $G_0$. Each of these structures can be associated with 
one of the peaks of the global conductance histogram of Fig. \ref{hist}(c), and to the evolution steps of nanowires along this 
direction.

\subsection{[110] Gold NW's}

\begin{figure}[h]
\includegraphics[width=6.5cm]{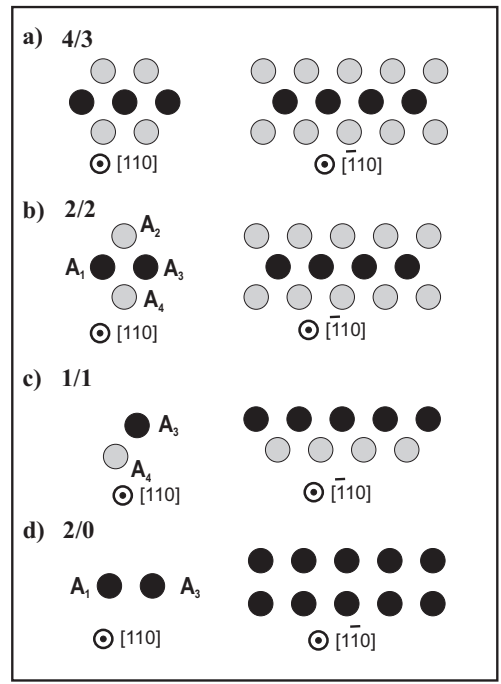}
\caption{Cross-section (left) and side view (right) of the schematic representation of possible rod-like [110] NW's at various stages of the
thinning process. The atomic arrangements are labeled from the number of atoms forming the alternate stacking planes generating the
rods: (a) 4/3, (b) 2/2, (c) 1/1 and (d) 2/0. See the text for a definition of the notation.} \label{110}  
\end{figure}

The HRTEM results show that compared to the other two kinds of nanowires, the [110] NW's are rather brittle, {\it i. e.}, they break 
when they are two-to-three atoms thick without forming atomic contacts. Figures \ref{110}(a-d) show the proposed 
structure of rod-like [110] NW's as they decrease in diameter, by loosing atomic planes\cite{Varlei1}. Again, the notation used in naming 
these structures comes from the number of atoms contained in each plane forming the NW (black and gray circles). However, in this case, 
due to their large aspect ratio, we show only the rod-like part and leave out the apexes. The thickest NW 
(Fig. \ref{110}(a)), labeled 4/3, is formed by two families of low-energy (111) facets and two high-energy (100) ones. By removing 3 lines 
of atoms from the previous structure we reduce it to the 2/2 NW. Finally we consider two possible scenarios: the 2/2 NW is subtracted of 
one of the (111) planes (atoms A$_1$ and A$_2$), giving rise to the alternate 2-atom chain, 1/1, shown in Fig. \ref{110}(c)); the 
other possibility is the elimination of the central plane of atoms (atoms A$_2$ and A$_4$), generating the parallel 2-atom chain 2/0
(Fig. \ref{110}(d)).

\begin{figure}[h]
\includegraphics[width=8.5 cm,clip=true]{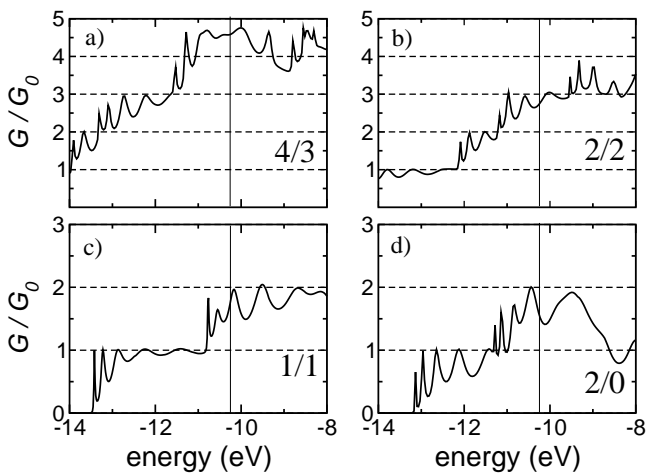}  
\caption{Calculations of the quantum conductance $G$ as a function of the electron energy for various 
rod-like NW's along the [110] direction: (a) 4/3, (b) 2/2, (c) 1/1, (d) 2/0 structures.
The vertical line indicates the $E_F$.} 
\label{theo-110} 
\end{figure}  

The calculated conductance for each of the [110] NW's presented in Fig. \ref{110} is shown in Fig. \ref{theo-110}. We can see that the 
conductance curves presented show oscillations that arise from interference effects inside the contacts. This oscillations are sensitive 
to the atomic positions so we must consider an average around the value E$_F$. The calculated 
conductance for the 4/3 structure, as presented in Fig. \ref{theo-110}(a), is $G(E_F)$ $>$ 4.5 $G_0$. Meanwhile, for the 2/2 case it is close 
to 2.8 $G_0$. The last two NW's structures, 1/1 and 2/0, show essentially the same conductance pattern, $G(E_F)$ $\sim$ 2 $G_0$. The last HRTEM 
image - within the time resolution considered - for NW's formed along the [110] direction is the 2/2 structure,
so it is not possible to resolve the final structure. However, the conductance histogram of Fig. \ref{hist}(d) shows that the last 
plateau is approximately 1.8 $G_0$, indicating that at least one of these structures must be present in UHV-MCBJ experiments. 

We might consider that the elongation process will increase the interplanar distance in the 2/2 structure and bring atoms A$_2$ and A$_4$ 
together until they reach first neighbor distances. Nonetheless the orientation of the stacking is changed to the [100] direction. As it
happened for the 732237 structure of section \ref{[111]}, the conductance calculated for this case shows slight decrease compared to the 
perfect structure (from 2 to 1.5 $G_0$).

In the light of the discussion above, we can model the atomic rearrangements in this direction by removing lines of atoms, starting at the
4/3 structure, which has a conductance $G(E_F)$ $>$ 4.5 $G_0$, down to the 2/2 that can be associated, aided by the calculations, to the peak 
at 2.8 $G_0$ on Fig. \ref{hist}(d). Finally, the peak at 2.0 $G_0$ is related to either the 1/1 or 2/0 structures.

\section{Effect of carbon impurities}\label{discuss2}

\begin{figure}[h]
\includegraphics[width=6.5cm]{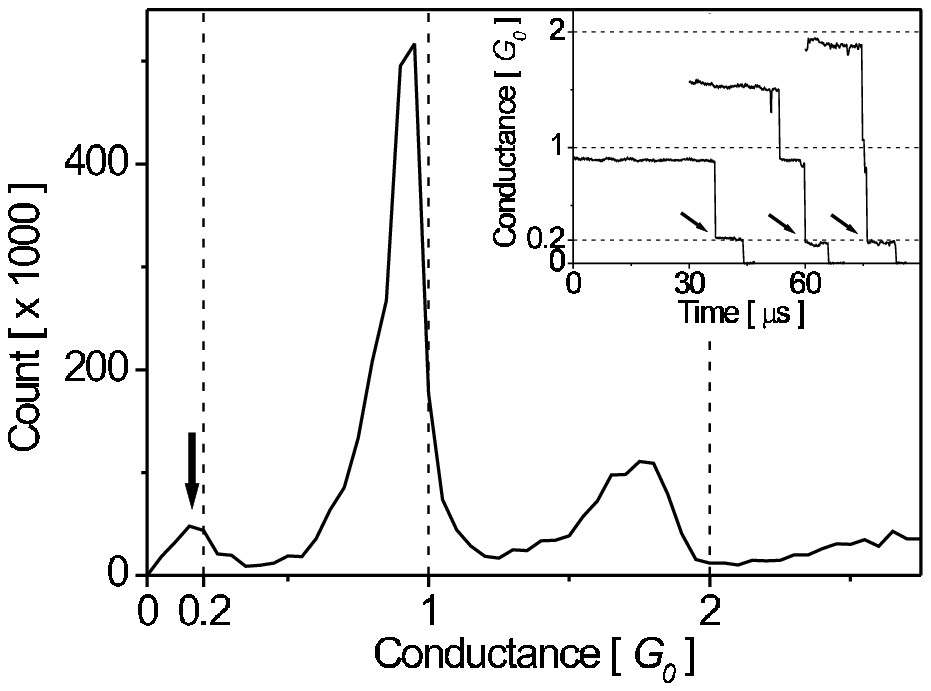}  
\caption{Conductance histogram of a set of 500 curves measured using the UHV-MCBJ after several hours of experiment. The arrow shows the
appearance of a peak between 0.1 and 0.2 $G_0$ which is not present in the first hours of experiment. Inset: Typical 
conductance curves; note the existence of plateaus at $G$ $\sim$ 0.2 $G_0$}\label{imp-exp} 
\end{figure}
 
The statistical correlation between the occurrence of conductance curve profiles and the multiplicity of zone axis has been hitherto 
described as a property of the Au nanowires.\cite{Varlei1} However, the accordance is only observed during the first hours of measurement 
after the macroscopic wire was fractured inside the MCBJ,
while the gold surfaces are  free of impurities; passed this initial period, contamination occurs even in UHV conditions. That 
is then evidenced by the appearance of conductance plateaus at values ranging approximately from 0.1 to 0.2 $G_0$ and 
whose occurrence increases with time.\cite{Varlei1} Similar observations have been reported for NW's seen using UHV scanning tunneling 
microscopy.\cite{RefOlesen1} The global
histogram and typical experimental conductance curves illustrating this effect are shown in Fig. \ref{imp-exp}. 
Likewise, another indication that the MCBJ might be contaminated  is the fact that the 1 $G_0$ plateaus become exceedingly stable and 
long, suggesting the formation of entire nanowires made of impurity atoms. One of the most probable contaminants is carbon. {\it Ab initio}
calculations\cite{Douglas} have shown that the carbon atom naturally occupies a position between two gold atoms in ATC's. Since the 
presence of carbon cannot be detected by HRTEM images because of its low atomic number,\cite{koizumi} the effect characterized by plateaus
at 0.1--0.2 $G_0$ (Fig. \ref{imp-exp}) is therefore the only experimental evidence of its presence. 

\begin{figure}[h]
\includegraphics[width=6cm,clip=true]{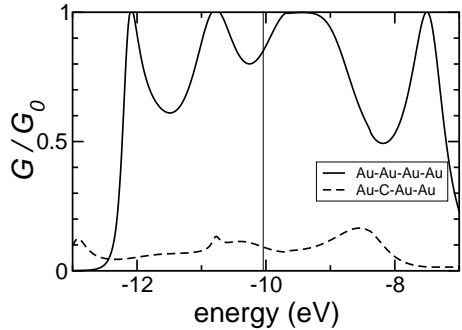}
\caption{Theoretical results for the conductance of ATC's as a function of the energy. The solid line corresponds to a 
pure gold ATC (interatomic distance 2.88\AA). The dashed line shows the effect of carbon impurities in the 
ATC, as depicted in the figure. The Au-C interatomic distance is fixed at 1.8 {\AA}, obtained from {\it ab initio} calculations. 
\cite{Douglas}} 
\label{imp-theo} 
\end{figure}

We have used the formalism of section \ref{theory} to address this question. For this purpose, consider the following gold ATC's in the 
[100] direction: a chain of 2 suspended gold atoms between two [100] tips (Au-Au-Au-Au, four Au atoms forming the narrowest region), as 
well as the contaminated Au-C-Au-Au chain. The assumed distance separating C and Au atoms is 1.8\AA,\cite{Douglas} whereas the Au-Au 
distance is taken to be 2.88\AA. Calculations of the conductance produced for the 2 cases are illustrated in Fig. \ref{imp-theo}. It can 
be verified that the presence of C atoms in the chain is responsible for decreasing the conductance of the nanocontact considerably from
1 to $\sim$ 0.1 $G_0$, in very good agreement with the experimental results and, moreover, corroborating the assumption that carbon can act
as a contaminant of Au NW's after a few hours of measurement.
 
\section{Conclusion}\label{conclu}
 
In summary, we have studied experimentally and theoretically the elongation of gold NW's, both from the structural evolution and the 
quantum conductance behavior. From the experimental point of view, we have used two independent set-ups: an UHV-MCBJ operating at room 
temperature for conductance measurements and HRTEM for real-time imaging of the NW atomic rearrangements. 

Gold NW's adopt only three kinds
of atomic structures where the elongation direction is parallel to one of the [100], [111] and [110] axis. For each one of these NW 
types, we have derived the three dimensional structures during different stages of the elongation process, by combining HRTEM images and 
the Wulff method. Subsequently conductance calculations, based on the extended H\"uckel theory and the Landauer formalism, were applied to
the deduced atomic structures, showing an excellent accordance with the conductance measurements. This approach has provided a consistent 
correlation between the structural evolution of the three kinds of Au NW's and the conductance behavior in the 0--3 $G_0$ conductance 
range. 

The EHT, although rather simple, allows us to take into account the atomic arrangement of the nanocontacs and, also it includes the 
possibility of considering  atoms of different species. As an example of its versatility, we have considered carbon impurities on 
suspended atom chains; the results indicate that they act as a contaminant causing the conductance plateaus to decrease from 1 to 0.1 
$G_0$. This result agrees extremely well with our conductance data of contaminated Au NW's and, also those from Olesen 
{\it et al}.\cite{RefOlesen1}

We have, therefore, thoroughly analyzed the relationship between the structural modifications involved in the elongation of Au nanowires 
and their transport behavior. The main conclusion that must be drawn from this work is the determinant role of an atomistic description 
when analyzing the conductance properties of nanoscopic metallic wires. 

\section{Acknowledgments} 
 
The authors are grateful to FAPESP, CNPq and LNLS for financial support.   
 
\section{Appendix} 
\label{appendix}

This appendix presents the details of the theory, circumvented in section \ref{theory}. When projected onto the basis states 
$\langle \xi'n'|$ the equation $H |\Psi \rangle = E |\Psi \rangle$ yields the following set of coupled equations for the set of 
coefficients $\{r_{\alpha\beta},t_{\alpha,\beta},C_{\phi,\alpha}\}$:
\begin{widetext}
\begin{eqnarray} 
\sum_{n<0} \sum_{\alpha} r_{\alpha\beta} e^{-in\theta_{\alpha}} A^{\xi'\alpha}_{n'n} 
+ \sum_{\phi} C_{\phi\beta} A_{n'0}^{\xi'\phi}  
+ \sum_{n>0} \sum_{\alpha} t_{\alpha\beta} e^{in\theta_{\alpha}} A^{\xi'\alpha}_{n'n} =  
- \sum_{n<0} e^{in\theta_{\beta}} A^{\xi'\beta}_{n'n} 
\label{eq1} 
\end{eqnarray}
\end{widetext} 
with  
\begin{eqnarray} 
A^{\xi'\xi}_{n'n} &=& \langle \xi'n'| H - E |\xi n \rangle \\ 
&=& H^{\xi'\xi}_{n'n} - E S^{\xi'\xi}_{n'n} \ . 
\label{A} 
\end{eqnarray} 
The term $H^{\xi'\xi}_{n'n}$  renders the MO energies of a given unit cell whenever $\xi'=\xi$ and $n'=n$. It also accounts for the 
hoppingenergies between the nanocontact and the first unit cell of the leads, on both sides, that is, 
$\langle \phi'n'|H|\alpha n \rangle = W^{\phi'\alpha}_{0\pm1}$. The same idea applies to the overlap terms $S^{\xi'\xi}_{n'n}$. Within the
leads we simplify the treatment by taking into account only the coupling between MO's of the same type, which yields 
$\langle \alpha'n'|H| \alpha n \rangle = h\ \delta_{\alpha'\alpha} \delta_{n',n\pm1}$ and 
$\langle \alpha'n'|\alpha n \rangle = S_0\ \delta_{\alpha'\alpha} \delta_{n',n\pm1}$, with $h$ and $S_0$ being the same for all MO's. This 
is necessary for obtaining analytical expressions for the dispersion energies of the electrons in the leads, which allows us to consider
explicitly only the sites $n=-1,0,1$ in (\ref{eq1}). Thus the energy dispersion relation is 
\begin{equation}
E(\theta_{\alpha})=\frac{\epsilon_{\alpha}+2h\cos\left(\theta_{\alpha}\right)}
{1+2S_{0}\cos\left(\theta_{\alpha}\right)} 
\end{equation}
and the velocity of the electron mode $v_{\alpha} = dE_{\alpha}/dk_{\alpha}$, with $k_{\alpha} = \theta_{\alpha}/a$ and $a$ is 
the length of the unit cell. The parameters $h=4\ eV$ and $S_0=0.15$ were chosen to yield a bandwidth of approximately $8\ eV$,
which is found in bulk Au.

Finally we obtain for the representative sites : \\
\begin{widetext} 
$n'=-1$ 
\begin{eqnarray} 
\sum_{\alpha}\left\{(h -ES_0)e^{i2\theta_{\alpha}} + (\epsilon^{\cal{L}}_{\alpha} -E)e^{i\theta_{\alpha}} \right\}
\delta^{\alpha'}_{\alpha} \ r_{\alpha\beta} 
+ \sum_{\phi} ( W^{\alpha'\phi}_{-10} &-& ES^{\alpha'\phi}_{-10} )  C_{\phi\beta} 
\nonumber \\  
= - \left\{ (h - ES_0)e^{-i2\theta_{\beta}} + (\epsilon^{\cal{L}}_{\alpha} - E)e^{-i\theta_{\beta}} \right\} 
\delta^{\alpha}_{\beta}  \ , & & 
\end{eqnarray} 
$n'=0$ 
\begin{eqnarray} 
\sum_{\alpha} \left(W^{\phi'\alpha}_{0-1} - E S^{\phi'\alpha}_{0-1}\right)e^{i\theta_{\alpha}} \ r_{\alpha\beta}  
+ \sum_{\phi} ( \epsilon_{\phi} -E) \delta^{\phi'}_{\phi}  C_{\phi\beta} 
&+& \sum_{\alpha} \left(W^{\phi'\alpha}_{0\ 1} - E S^{\phi'\alpha}_{0\ 1}\right)e^{i\theta_{\alpha}}\ t_{\alpha\beta}
\nonumber \\ 
= - \left(W^{\phi'\beta}_{0-1} - E S^{\phi'\beta}_{0-1}\right)e^{-i\theta_{\beta}} \ \delta^{\alpha}_{\beta} \ , && 
\end{eqnarray} 
$n'=1$ 
\begin{eqnarray} 
\sum_{\alpha} \left\{ (h-ES_0)e^{i2\theta_{\alpha}} + (\epsilon^{\cal{R}}_{\alpha} -E)e^{i\theta_{\alpha}} \right\}
\delta^{\alpha'}_{\alpha} \ t_{\alpha\beta} 
+ \sum_{\phi} ( W^{\alpha'\phi}_{1\ 0} -E S^{\alpha'\phi}_{1\ 0} )  C_{\phi\beta}= 0
\end{eqnarray} 
\end{widetext}

\end{document}